\newcommand{\ppbar}{{p\bar{p}}}
\documentclass[xft_HCPproceedings]{revtex4}
\usepackage{graphicx}
\usepackage{fancyhdr}
\begin{document}

\title{{\small{Hadron Collider Physics Symposium (HCP2008),
Galena, Illinois, USA}}\\ 
\vspace{12pt}
The CDF L2 Track Trigger Upgrade} 

%

\author{D. Cox}
\affiliation{University of California - Davis}
%
%

\begin{abstract}
This proceedings describes the XFT stereo upgrade for the CDF Level 2 trigger system. Starting with the stereo finder boards, up to the XFT stereo track algorithm implementation in the Level 2 PC. This note will discuss the effectiveness of the Level 2 Stereo track algorithm at achieving reduced trigger rates with high efficiencies during high luminosity running.
\end{abstract}

\maketitle

\thispagestyle{fancy}


\section{INTRODUCTION} 
Since 2001 the CDF II detector has been collecting data used to carry out a rich physics program at the Fermilab Tevatron $\ppbar$ collider. The quality and purity of the data relies heavily on the high performance and efficiency of the CDF trigger system. The heart of the trigger for physics is the eXtremely Fast Tracker (XFT), a trigger track system used to identify charged tracks. Tracks are identified based on the data from sense wires in the central tracking chamber in time for the Level 1 trigger decision, and are extrapolated to the calorimeter and muon chambers to form electron and muon trigger candidates. 

The steady increase of the Tevatron instantaneous luminosity results in higher detector occupancy from multiple $\ppbar$ interactions. For the track trigger, this produces overlapping pattersn of hits identified as high momentum tracks and lead to the rapid growth in trigger rates. The control of trigger rates by raising thresholds or prescaling triggers results in signficicant loss in acceptance of important physics signatures with a negative impact on the CDF II physics program that demanded an upgrade of the original system. Without the upgrade the physics potential of CDF would be significantly compromised.

The track segments identified at Level 1 are delivered to the Level 2 trigger proccessors, where more available time allows 3D-track reconstruction improving the track trigger purity and allowing use of $\phi$ and cot $\theta$ information. Here we provide a brief overview of the XFT system and the Level 1 track trigger upgrade and a description of the Level 2 part of the XFT system.

\section{AXIAL XFT SYSTEM LEVEL 1 STEREO UPGRADE}
The CDF II tracking chamber has sense wires arranged in 8 super-layers alternating between axial and $\pm2^{0}$ stereo. Twelve sense wires are grouped into drift cells. The innermost superlayer contains 168 stereo cells, while the outermost one has 480 axial drift cells. Eight adjacent cells are read out by TDC modules that digitize the arrival time of analog hit signals induced on each wire.

The XFT track processing consists of three stages: hit classification on each sense wire, track segment finding based on predefinied characteristic patterns of groups of hits from real tracks, and linking segments between super-layers to identify tracks.

It was forseen that the Tevatron would operate with 108 bunches of protons and anti-protons. Therefore each axial sense wire provides two bits of drift time information prompt or delayed hit, every 132 ns. In reality, the Tevatron operates with 36 bunches and a 396 ns bunch crossing interval. The stereo upgrade makes use of it by allowing three times more information to be transferred per wire. The mezzanine card (XTC2) reports hit times on stereo super-layers in 6 bins. The hit data is transferred to Finder modules every 16.5 ns.

The Finder modules identify whether hits from adjacent cells are consistent with predefined patterns corresponding to segments from tracks with $p_{T} > 1.5$ GeV/c. Each Finder module processes hit data and finds track segments in a slice of $\phi = 30^{0}$ at one of the super-layers. At Level 1, twelve bits of data are referred to as ``pixels'' are used to identify the azimuthal positions of track segments in the inner two axial super-layers (numerated SL 2 and 4) and six groups of two bits in the outer two axial superlayers (SL 6 and 8) and the outer three stereo superlayers (SL 3,5 and 7). The two bits at each azimuthal position provide information about positive or negative curvature of the track. Both bits are on for high-momentum tracks with $p_{T} > 8$ GeV/c. Higher pixel granularity in the Stereo Finders is also made available to Level 2 and discussed in the next section. 

The axial track segment data is sent to Linker Modules which match groups of pixels in the four axial super-layers to predefined patterns corresponding to valid tracks. The stereo pixel data from the Stereo Finders is sent to Stereo Linker Association Modules (SLAM) via a fiber transmitter mezzanine board. The SLAM modules also receive the list of axial tracks found by the Linker Modules via the crate backplane. The SLAM modules perform the association of stereo pixels with axial tracks by exploiting the correlation between the azimuthal position of an axial track and the distances to the associated pixels foudn in the stereo super-layers. Because of the $\pm2^{0}$ stereo angle on alternating stereo super-layers, stereo pixels are alternatively displaced into the positive and negative direction depending on the polar angle of the track.

If the axial track matches to at least one predefined pattern of stereo pixels the track is considered to be stereo confirmed, and this bit of the information along with the axial properties of the track ($p_{T}$ and $\phi_{0}$) is driven to the XTRP system which extrapolates tracks to outer sub-detectors in the plane transverse to the beam direction.

The Level 1 track trigger upgrade improves the fake track rejection by a factor of 3 to 5 and is 97\% efficient.

\section{LEVEL 2 STEREO UPGRADE}
The Level 2 stereo upgrade is an extension of the Level 1 track trigger upgrade. It introduces a new data stream to the Level 2 trigger processors and thus significantly enhances the triggering capabilities at CDF. This data stream consists of the stereo track segements list with higher $\phi$-pixel and track slope granularity than is available to Level 1.

\begin{figure}
  \begin{center}
\includegraphics[width=15cm]{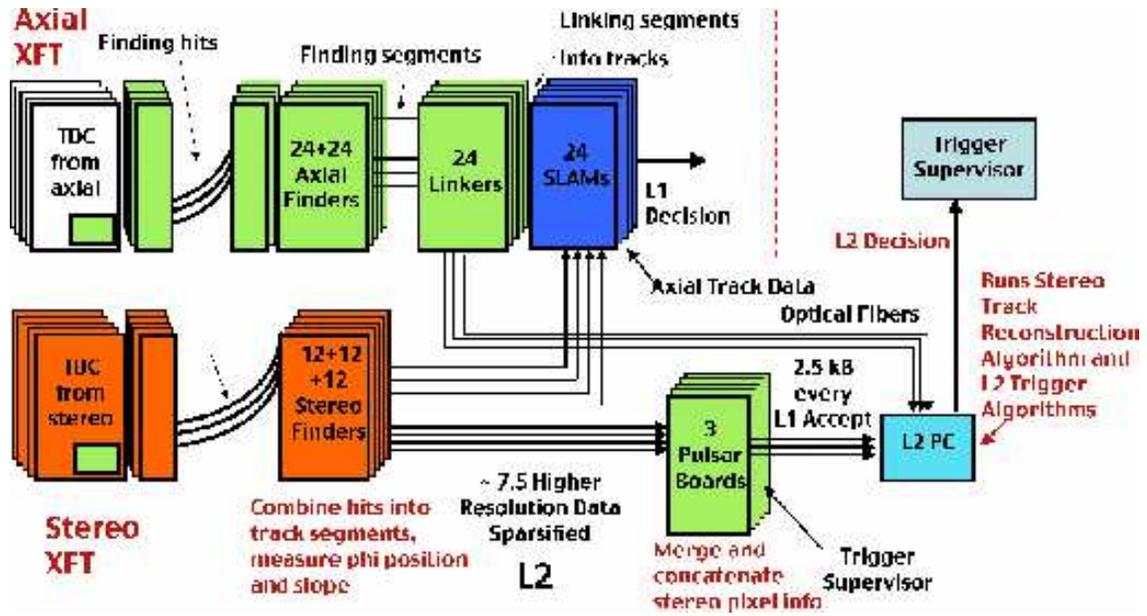}
  \end{center}
  \caption{Configuration of the eXtremely Fast Tracker (XFT) system.}
  \label{fig:configuration}
\end{figure}

The Level 2 system configurtion together with the axial and the Level 1 stereo part of the system is shown in Figure \ref{fig:configuration}

\subsection{Level 2 Stereo Track Segments Finding and Sparsification}
The XFT stereo track finding mechanism is based on analyzing combinations of hits. Each combination of 6 time bins for each of the 12 sense wires, not neccessarily within the same cell, defines a pattern. For each pattern, a pixel corresponding to a $\phi$ position of the track crossing a given super-layer and the slope of the track is determined based on the extrapolation of tracks with random curvature, $p_{T} > 1.5$ GeV/c, through the tracking chamber geometry described with a drift time model. In general, the number of patterns corresponding to the same pixel and slope location can be large, and certain simplifications in the pattern generation are emplaced so as not to exceed the limitations imposed by the chip's resources.

The Stereo Finder modules utilize two Altera Stratix 2EP2S60 \cite{altera} FPGA's, and the genrated patterns are used as the input for the segment finding firmware. Two Finder FPGAs divide the hit data into two categoies, each one processing 18 or 10 cells depending on the superlayer, SL7 and SL3 respectively. The Stereo Finder processing the data from superlayer 5 has an asymmetric design with one FPGA handling 18 and the other one handling 10 cells of the information.

Every 16.5 ns for 18 consecutive clocks the pattern in the Finder FPGA produces 90 bit words per each cell. Each bit of this information corresponds to a Level 2 stereo pixel with 18 $\phi$ and 5 slope possibilities per cell. These 90 Level 2 stereo pixels get 'OR'ed down to 12 pixels with 6 $\phi$ and 2 slope bins which are transmitted to the SLAM modules to be used in the Level 1 trigger. If a Level 1 Accept is received the 7.5 times higher granularity Level 2 stereo are stored into RAM and then processed through the VHDL sparsifier state machine.

Six groups of pixels corresponding to Level 1 $\phi$ positions within a cell define the subcells and form the header section of the Level 2 output data stream. Each clock tick a bit of this information is analyzed and, if the subcell is fired, a 15 bit (3 $\phi$ \& 5 slope) word of data is retreived from RAM. The sparsified data from two Finder FPGA's is transmitted to a Level 2-Pulsar FPGA operating at a different clock, which reformats the data and transfers it to a 4-channel fiber transmitter mezzanine board, designed according to the Common Mezzanine Card (CMC) \cite{Anikeev} standard. The fiber transmitter uses TLK1501 \cite{tlk} devices to serialize the data into 16-bit words, which are transferred to Pulsar boards every 16 ns.

\subsection{Merging and Transmission of the Stereo Data}
PULSAR (stands for ``PULSer and Recorder'') \cite{PULSAR} is a general purpose 9U VME board, which interfaces different upstream detector subsystems, and was designed primarily as an upgrade path for the new CDF II Level 2 trigger system commissioned in the summer of 2005. Data from Stereo Finder modules are received by the 4-input optical receiver mezzanine boards developed to carry data in the XFT data format. These are the same boards used on the Stereo Finders to receive data from the TDCs. 

The Pulsar board utilizes three FPGAs (Altera APEX 20K400BC-652-1XV \cite{APEX}): two DataIO FPGAs and one Control FPGA. Both Data IOs FPGAs provide interfaces to two mezzanine cards each. For the XFT stereo upgrade application, the Pulsar board performs the merging of the segment list data from 12 Stereo Finders and re-formats it into an S-LINK 32-bit word standard packet \cite{SLINK}

The S-LINK data format allows communication between the Pulsar board and S-LINK-to-PCI interface Four Input Links for Atlas Readout (FILAR) card \cite{FILAR}. The FILAR is a high bandwidth S-LINK-to-PCI interface card developed in CERN and designed to have low PCI bus utilization with minimal host processor control. Each FILAR card performs an autonomous data reception from four S-LINK channels delivering data from various detector subsystems to the Level 2 decision node PC memory. The Level 2 decision node is a Dual Core AMD Opteron 290 2.8 GHz (4 cores), 2GB RAM, gentoo Linux machine that unpacks received data packets from various detector components and runs the trigger algorithms. The decision packet is further sent out via the S32PCI64 PCI interface card, a transmitter analog of the FILAR with one channel, to another Pulsar board that communicates the decision to the Trigger Supervisor.

In contray to the data received from most other detector subsystems that consist of already processed information such as a list of axial tracks, the XFT stereo segments data is ``raw'', and preceeding the trigger algorithms it needs to be proccessed by the decision node CPU. The XFT stereo segment data volume is therefore larger than any other data packet from other detector components. It generally ranges from 1.5 to 3 kB per event, which accounts form more than ~50\% of the total data volume transferred through the PCI bus. To ensure that this enlarged data volume does not cause problems to the Level 2 XFT system the XFT Pulsars have abort and truncation functionality. The abort functionality reduces average demand by not transmitting any XFT stereo data on events where it will not be needed. The truncation functionality terminates stereo data transmission after a settable number of data words, cutting off any events with unusually high data volumes. Both aborted and truncated events are treated as automatic stereo acceptances by any triggers that call upon the stereo reconstruction.

\subsection{Stereo Track Reconstruction}
\begin{figure}
  \begin{center}
\includegraphics[width=15cm]{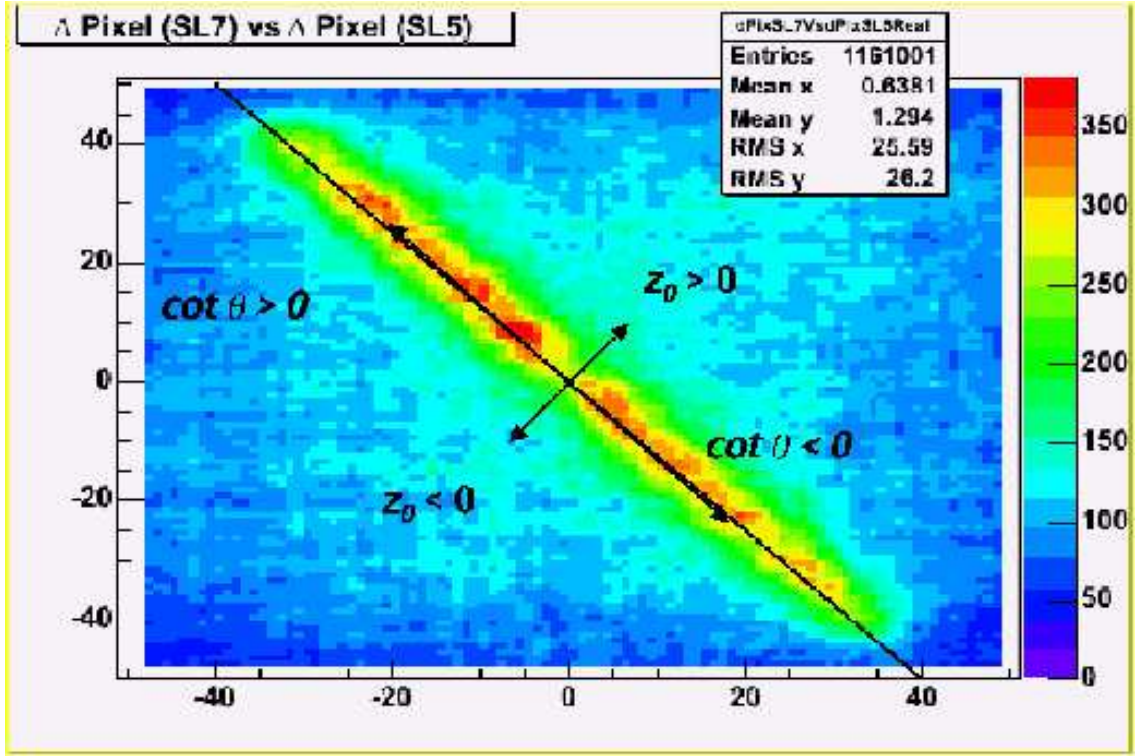}
  \end{center}
  \caption{Displacements of the azimuthal positions of stereo pixels in the outer two stereo super-layers. The pixels from real trracks are aligned while pixels from fakes uniformly populate the plane. Stereo track properties ($z_0$ and cote $\theta$) can be computed based on the measured displacements of pixels.}
  \label{fig:delpixel}
\end{figure}

Because of the high volume of the XFT stereo data packet and the highly sparsified format, it is inefficient to unpack it fully, and therefore the unpacking is performed on demand only for regions of interest, i.e. near the position of a traversing axia, stereo-confirmed by Level 1, track. The stereo track reconstruction algorithim consists of the following steps. First, the axial track is extrapolated to each of the outer three stereo superlayers. Next, at each stereo superlayer the track segments corresponding to $\pm$3 cells centered near the extrapolated $\phi$ position of the track are unpacked. Due to alternating $\pm$2$^O$ stereo angel, pixels corresponding to the real track are alternatively displaced as is shown in Figure \ref{fig:delpixel}. The correlation between displacements at various superlayers is exploited in the Level 1 trigger, while at Level 2 higher pixel granularity also allows the determination of the stereo properties of the track that can be expressed in terms of $z_0$ and cot $\theta$.

The stereo track reconstruction performance deteriorates as the number of fake segments increases with the instantaneous luminosity. There are a total of 1620 bits of information (6 cells $\times$ 18 $\phi$ pixels $\times$ 3 superlayers $\times$ 5 slopes) to be analyzed. At the next stage, the slopes of the extrapolated track for each super-layer are obtained, and the pixels with slopes inconsisten to the ones of the track are ignored, and then pixels corresponding to the same $\phi$ position are 'OR'ed between slopes. This operation decreases the amount of bit information by a factor of 5 by filtering a large number of fake segments.

Next, the fired pixels across different stereo layers are combined into triplets. To decrease the combinatorics in case of the two or three adjacent pixels fired, the mean of the cluster of pixels is used in the triplet. Only aligned triplets are considered that extrapolate to the luminous region of the detector $|z_0|$ less than 60 cm). The trigger utilizing the stereo reconstruction determines which of these surviving triplet combinations are examined.

\begin{figure}
  \begin{center}
\includegraphics[width=8cm]{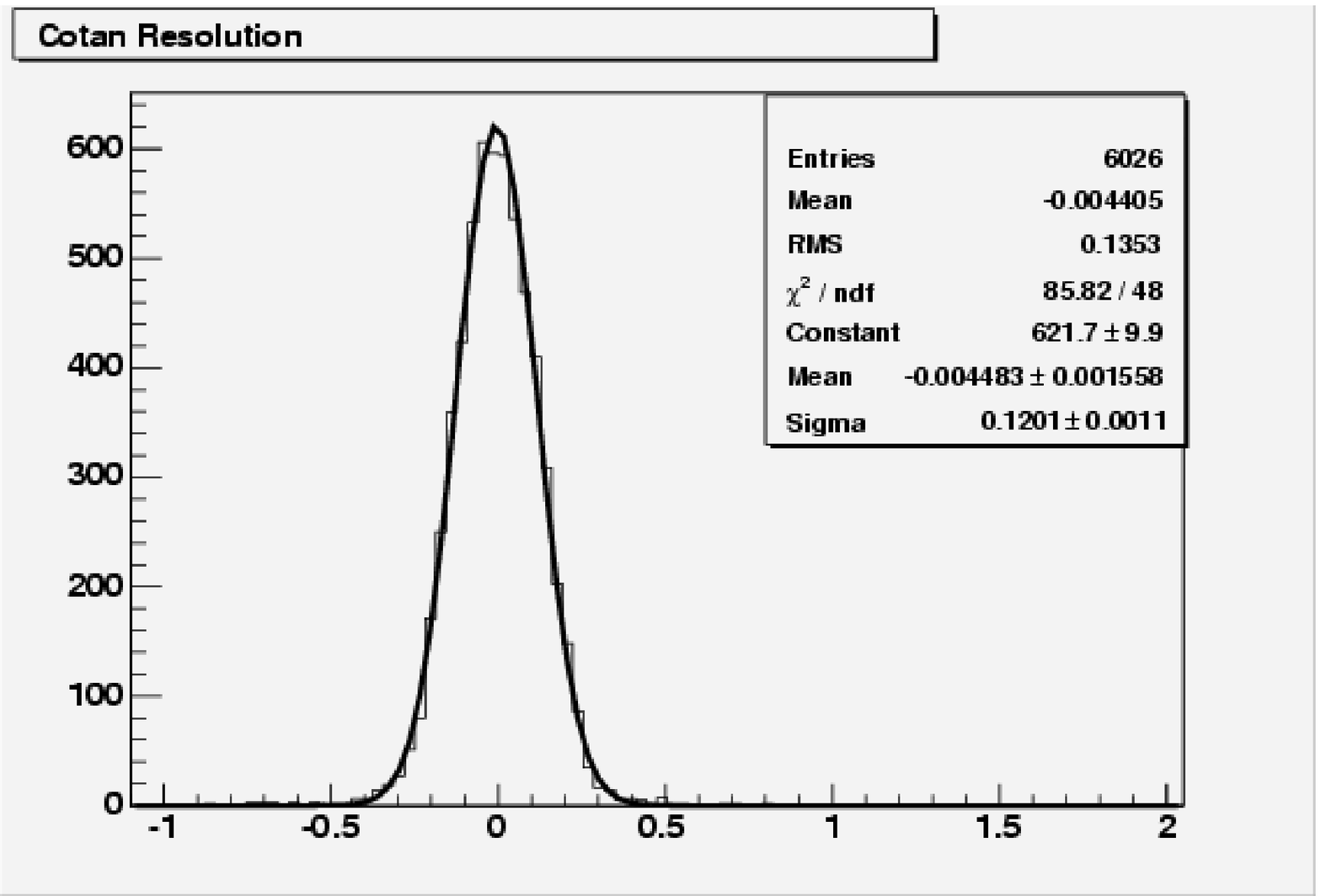}
\includegraphics[width=8cm]{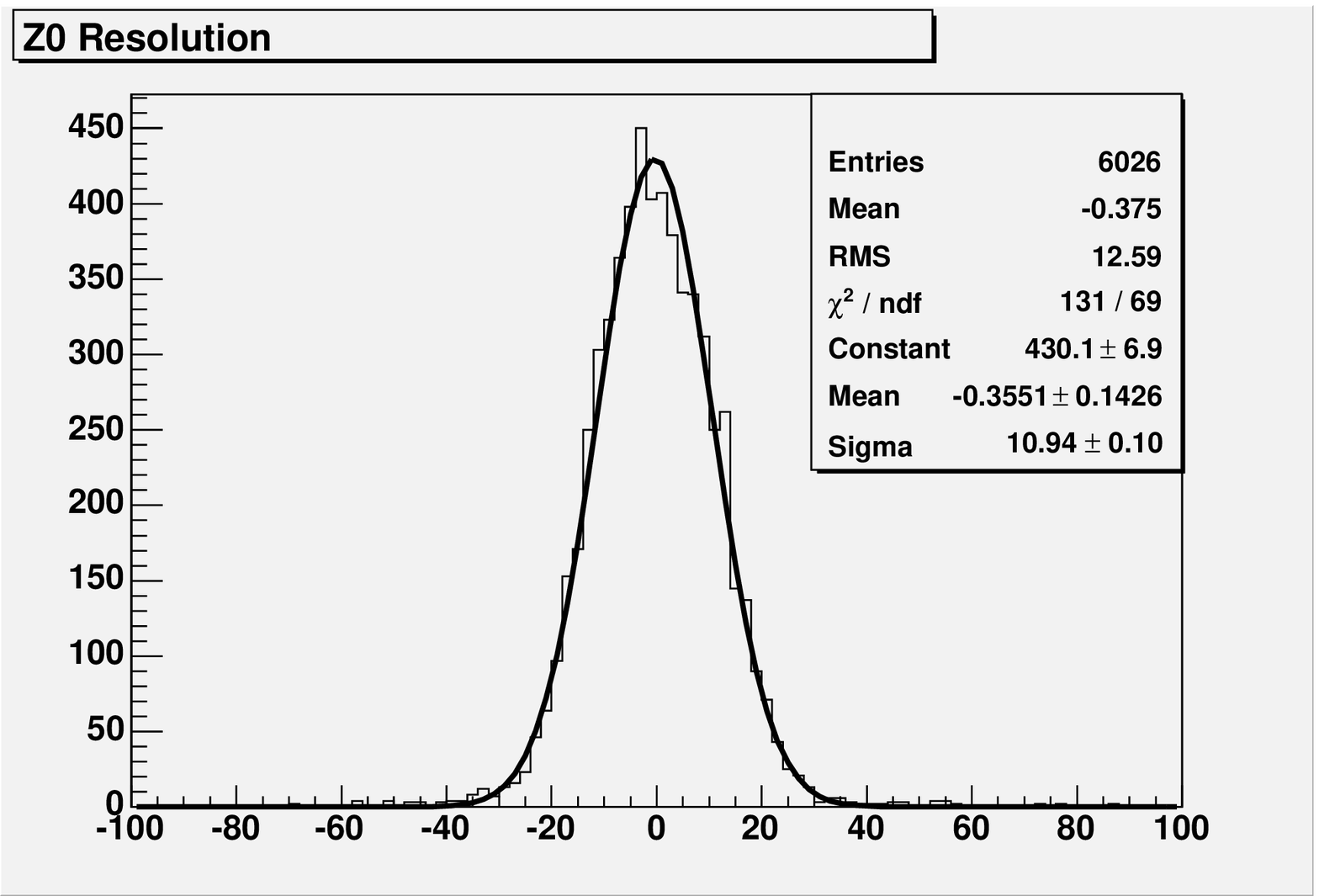}
  \end{center}
  \caption{The cot $\theta$ and $z_0$ resolutions of the Level 2 stereo tracks}
  \label{fig:resolutions}
\end{figure}

The procedure described above provides measurements of $z_0$ and cote $\theta$ at the Level 2 with resolutions of 11 cm and 0.13 rad respectively (see Figure \ref{fig:resolutions}). It is important to note that the resolution distributions are close to gaussian thus assuring the high efficiency of stereo trak matching to detector components.

\begin{figure}
  \begin{center}
\includegraphics[width=15cm]{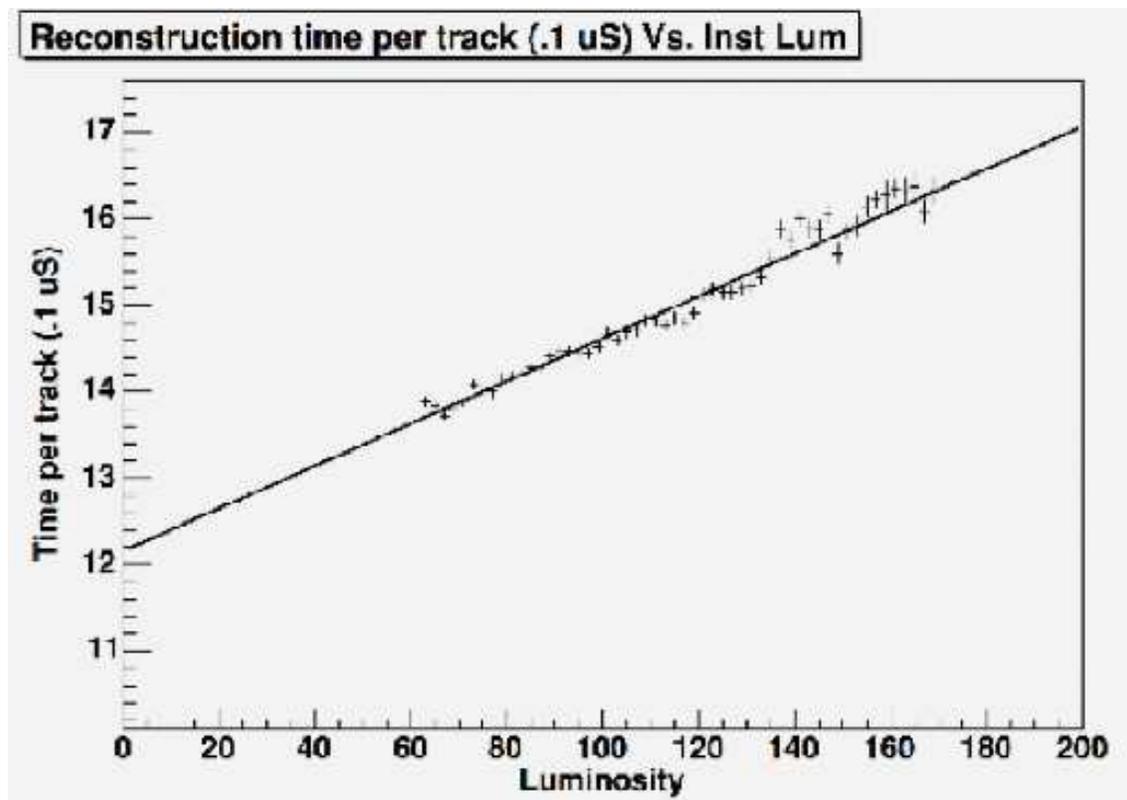}
  \end{center}
  \caption{CPU time required for stereo track reconstruction.}
  \label{fig:reconstruction}
\end{figure}

All of the steps of the stereo track reconstruction algorithim are optimized and reduced to a sequence of bit-wise operations. Nevertheless, it is a rather CPU intensive process. Figure \ref{fig:reconstruction} shows the time spent to reconstruct a track as a function of instananeous luminosity. To reduce the effect on the Level 2 latency of the system, stereo tracking is performed on demand, i.e. only if the track passes trigger requirements applied to its axial quantities.

\subsection{Trigger Rate Reduction}
The CDF II detector is operated at high instantaneous luminosities with the goal of maximizing trigger acceptance of high $p_T$ physics processes. At peak luminosity, the primary limitation is the Level 2 bandwidth of 900 Hz.

\begin{figure}
  \begin{center}
\includegraphics[width=15cm]{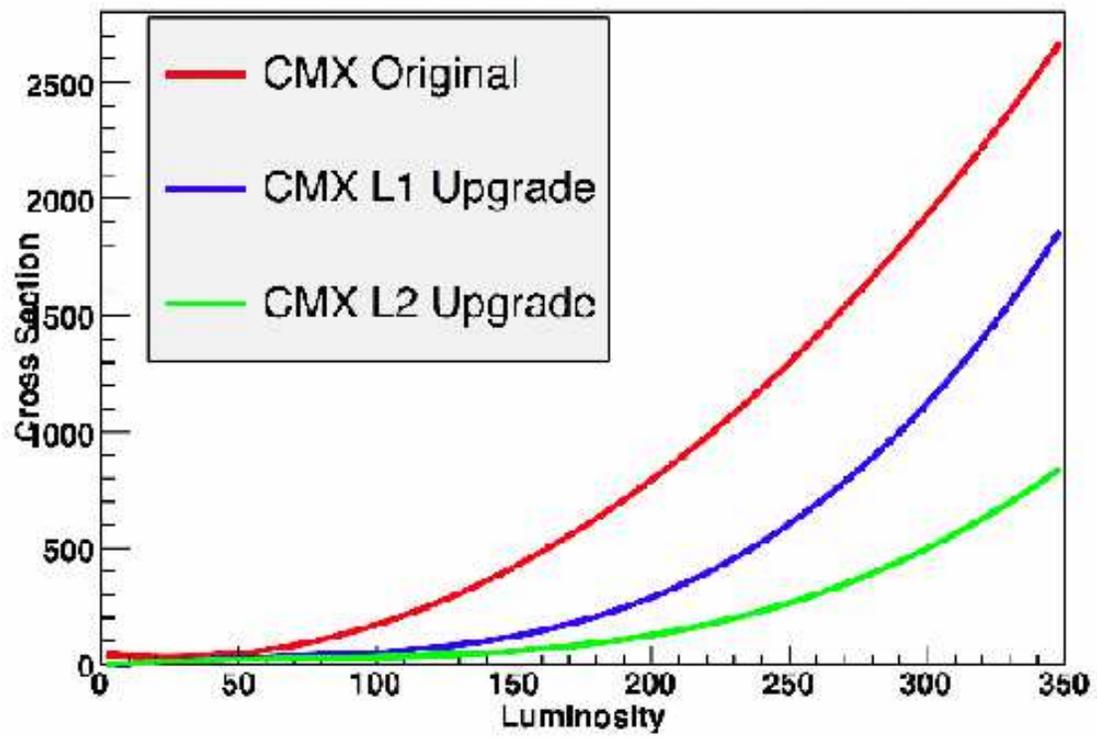}
  \end{center}
  \caption{Trigger cross sections for Level 2 CMX muon trigger.}
  \label{fig:performance}
\end{figure}

The Level 1 track trigger upgrade improves the track purity and reduces a trigger rates by a factor between 3 and 5 as evidenced from the CMX trigger before and after the upgrade (see Figure \ref{fig:performance}). The stereo tracking at Level 2 takes advantage of more information and provides reduction by a a factor ranging from 2 to 4.

The Level 1 upgrade was completed in Fall 2006 and has been used for collecting data since then. The Level 2 part of the upgrade was comissioned near the end of 2007 and has been used for collecting data since then.

\section{CONCLUSIONS}

The 3D track trigger upgrade of the CDF II detector has been described. The upgrade introduces new capabilities of 3-D track reconstruction at Level 2, considerably improves the performance and efficiency of the CDF II trigger system, especially at high instantaneous Tevatron luminosities and enhances the physics potential of the CDF Run II detector.

\begin{thebibliography}{9}
\bibitem{altera}
Altera Corporation, ``Stratix II Device Handbook'', August 2006.
\bibitem{Anikeev}
K. Anikeev, $et al.$, IEEE Trans. Nucl. Sci. TNS-00560-2004.
\bibitem{tlk}
Texas Instruments, ``TLK1501: 0.6 to 1.5 GBPS Transceiver,'' Data sheet, 2004.
\bibitem{PULSAR}
K. Anikeev, $et al.$, FERMILAB-PUB-06-400-E, 2006. 6pp.
\bibitem{APEX}
ALTERA Pub., ``APEX 20k Programmable Logic Device Family'', Data Sheet v. 5.1, 2004
\bibitem{SLINK}
E. van der Bij, $et al.$, (1997, Sep.) ``S-Link, a Data Link Interface Specification for the LHC Era.'' Presented at 10th IEEE Real Time Conference. http://hsi.web.cern.ch/HSI/s-link
\bibitem{FILAR}
W.Iwanski, $et al.$, (2001, Aug.) ``Designing an S-LINK to PCI Interface using an IP core.'' Presented at 12th IEEE-NPSS Real Time Conference. http://hsi.web.cern.ch/HSI/s-link/devices/filar
\end{thebibliography}
\end{document}